\documentclass[aps,prl, amsmath,amssymb,twocolumn,superscriptaddress,reprint]{revtex4-2}

\usepackage[utf8]{inputenc}
\usepackage{graphicx}
\usepackage{placeins}
\usepackage{float}
\usepackage{subfigure}
\usepackage{amssymb}
\usepackage{soul}

\usepackage[table,xcdraw]{xcolor}
\usepackage{makecell}

\usepackage{natbib}
\usepackage{hyperref}
\usepackage{amsmath}
\usepackage{siunitx}
\usepackage{xcolor}
\usepackage{csquotes}
\usepackage{nicefrac}

\DeclareUnicodeCharacter{0301}{\textvisiblespace}
\DeclareUnicodeCharacter{2212}{\textendash}
\hypersetup{
    colorlinks=true,
    linkcolor=blue,
    filecolor=magenta,      
    urlcolor=blue,
    citecolor=blue,
}

\begin{document}

\title{Signatures of canted antiferromagnetism in infinite-layer nickelates studied by x-ray magnetic dichroism}

\author{G. Krieger}
\affiliation{Université de Strasbourg, CNRS, IPCMS UMR 7504, F-67034 Strasbourg, France}
\altaffiliation[present address:]{Eindhoven University of Technology, P.O. Box 513, 5600 MB Eindhoven,The Netherlands}
\author{H. Sahib}
\affiliation{Université de Strasbourg, CNRS, IPCMS UMR 7504, F-67034 Strasbourg, France}
\author{F. Rosa}
\affiliation{Dipartimento di Fisica, Politecnico di Milano, Piazza Leonardo da Vinci 32, I-20133 Milano, Italy}
\author{M. Rath}
\affiliation{CNR-SPIN Complesso di Monte S. Angelo, via Cinthia - I-80126 Napoli, Italy}
\author{Y. Chen}
\affiliation{CNR-SPIN Complesso di Monte S. Angelo, via Cinthia - I-80126 Napoli, Italy}
\author{A. Raji}
\affiliation{Laboratoire de Physique des Solides, CNRS, Université Paris-Saclay, 91405 Orsay, France}
\affiliation{Synchrotron SOLEIL, L’Orme des Merisiers, BP 48 St Aubin, 91192 Gif sur Yvette, France}
\author{P.V.B. Pinho}
\affiliation{ESRF, The European Synchrotron, 71 Avenue des Martyrs, F-38043 Grenoble, France}
\author{C. Lefevre}
\affiliation{Université de Strasbourg, CNRS, IPCMS UMR 7504, F-67034 Strasbourg, France}
\author{G. Ghiringhelli}
\affiliation{Dipartimento di Fisica, Politecnico di Milano, Piazza Leonardo da Vinci 32, I-20133 Milano, Italy}
\author{A. Gloter}
\affiliation{Laboratoire de Physique des Solides, CNRS, Université Paris-Saclay, 91405 Orsay, France}
\author{N. Viart}
\affiliation{Université de Strasbourg, CNRS, IPCMS UMR 7504, F-67034 Strasbourg, France}
\author{M. Salluzzo}
\affiliation{CNR-SPIN Complesso di Monte S. Angelo, via Cinthia - I-80126 Napoli, Italy}
\author{D. Preziosi}\email[]{daniele.preziosi@ipcms.unistra.fr}
\affiliation{Université de Strasbourg, CNRS, IPCMS UMR 7504, F-67034 Strasbourg, France}
\begin{abstract}
We report an experimental study of the magnetic properties of infinite-layer Nd$_{1-x}$Sr$_x$NiO$_2$ ($x$ = 0 and 0.2) thin films by x-ray magnetic circular dichroism (XMCD) at Ni $L_{3,2}$- and Nd $M_{5,4}$-edges. We show the presence of a field induced out-of-plane Ni$^{1+}$ spin-moment, reaching values of 0.25 $\mu_{B}/Ni$ at 9\,T in the case of superconducting Nd$_{0.8}$Sr$_{0.2}$NiO$_2$. 
The magnetic field and temperature dependencies of the Ni $L_{3,2}$ XMCD data can be explained by an out-of-plane canting of in-plane anti-ferromagnetically correlated Ni$^{1+}$ spins. The canting is most likely attributed to the symmetry lowering of the NiO$_{2}$ planes, observed via four-dimensional scanning transmission electron microscopy, which can trigger a Dzyaloshinskii–Moriya interaction among the Ni$^{1+}$ spins. 
\end{abstract}
\maketitle

The recent discovery of superconductivity in infinite-layer (IL) nickelate thin films \cite{Li2019} is shedding new light on the so-long-sought path towards the understanding of high-temperature superconductivity. Most of the ongoing research is devoted at evidencing differences and similarities between cuprates and IL nickelates, sharing similar square-planar structure and $3d^9$-electron count \cite{Wang2024}.
An important question yet to be settled is the difference in the magnetic ground-state, in view of the apparent absence of long-range antiferromagnetic (AFM) order in undoped polycrystalline NdNiO$_2$ \cite{HAYWARD2003839,Li2020Absence} and LaNiO$_2$ \cite{Hayward1999, Ortiz2022, Zhao2021} samples. The results were initially attributed to local oxygen non-stoichiometry and Ni$^{2+}$-related defects, besides the possible presence of ferromagnetic impurities, which would explain also the absence of bulk superconductivity \cite{Li2020Absence}.
Theoretically, most of the studies suggest that the energetically-favored AFM order of the Ni $3d^9$ system is more unstable than in cuprates, in particular when considering the role of the rare-earth magnetism \cite{PickettAFM,Lechermann2021, Zhang2023, Sahinovic2023,Hepting2020}(plus ref. therein). Indeed, G-type and C-type AFM ground states are found to be very close in energy, and Sr-doping has been predicted to favor also a transition from G-AFM towards C-AFM \cite{Sahinovic2023}. 
Concerning IL nickelate thin films, only few studies explicitly addressing the magnetism are available \cite{Fowlie2022, LuMagnon,RossiMagnetism}. This is particularly relevant, as the electronic and magnetic properties of thin films can be different from bulk samples.
Muon-spin rotation/relaxation measurements on SrTiO$_3$-capped undoped and superconducting R$_{1-x}$Sr$_{x}$NiO$_{2}$ thin films ($R$= Nd, Pr and La), revealed a complex spin-dynamics, with a temperature dependence of the spin-susceptibility pointing to an intrinsic magnetic order attributed to the Ni-spin-sublattice, despite the magnetism stemming from the $R$ $4f$-states in some sample series \cite{Fowlie2022}. Noteworthy, all these studies, on bulk and thin film samples, could not explicitly address separately the magnetism of the Ni$^{1+}$ and $R$ sublattices, nor single out the contributions from unwanted impurities and from Ni-ions with a $2+$ valence state that could be present due to an incomplete oxygen reduction, as highlighted in recent works \cite{Raji2023,Parzyck2024}.
On the other hand, resonant inelastic x-ray scattering (RIXS) experiments in epitaxial films show dispersing magnetic-excitations (magnons) within the NiO$_2$ planes \cite{LuMagnon,kriegerPRL,RossiMagnetism}, analogous to those of doped cuprates, with a next-neighbour AFM exchange coupling only half the value found in cuprates \cite{Braicovich2009}. Sr-doping, while triggering superconductivity, gives rise to a flattening-out of the magnetic-excitations \cite{LuMagnon,Higashi2021,rosa2024spin}. The observation of magnons in nickelate thin films points at an important role played by AFM correlations like in cuprates, where superconductivity emerges from an AFM ground state. 

Here, we report a study of the intrinsic magnetic properties of the Ni$^{1+}$-spin-sublattice in Nd-based IL thin films, by performing element(and valence)-specific x-ray magnetic circular dichroism (XMCD) experiments at the Ni $L_{3,2}$- and Nd $M_{5,4}$-edges as a function of magnetic field (H) and temperatures (T) on a series of  Nd$_{1-x}$Sr$_x$NiO$_2$ ($x$ = 0 and 0.2) samples. We find evidences of AFM-correlations in the Ni$^{1+}$-spin-sublattice, but at the same time a large, field induced, Ni$^{1+}$ out-of-plane spin-moment, a factor 10 larger than the canted Cu$^{2+}$ spin-moment determined by XMCD in cuprates \cite{DeLuca2010}. The results are explained by a canted antiferromagnetism of the Ni$^{1+}$ spin induced by a symmetry lowering of the NiO$_2$ planes, as observed by four-dimensional scanning transmission electron microscopy.
\begin{figure*}[t!]
    \centering
    \includegraphics[width=0.9\textwidth]{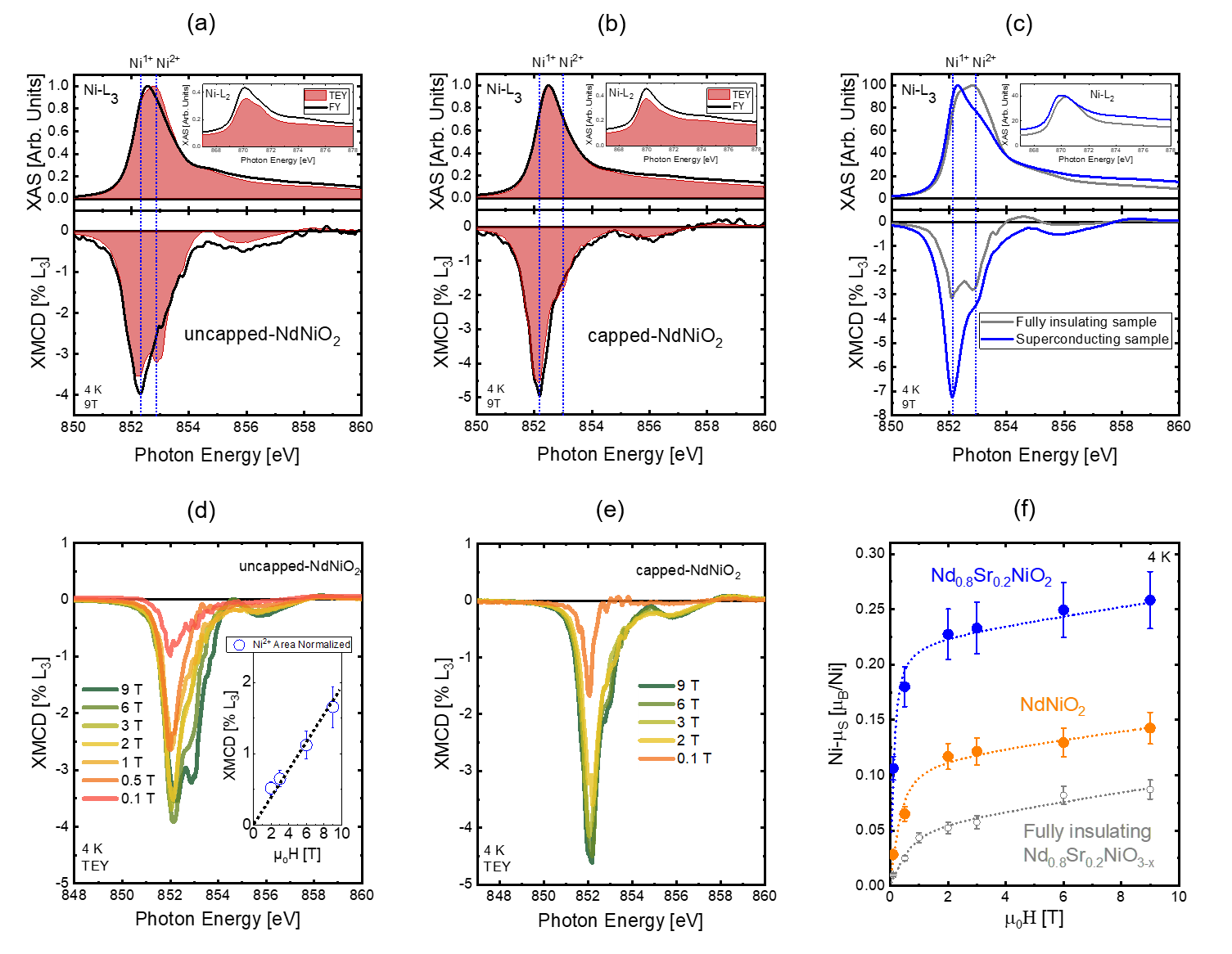}
    \caption{XAS (top part) and XMCD (bottom part) of (a) uncapped, (b) capped NdNiO$_2$ and (c) capped Nd$_{0.8}$Sr$_{0.2}$NiO$_2$ thin films acquired in TEY and FY modes at 9\,T and 4\,K. In (c) we compare data on superconducting Nd$_{0.8}$Sr$_{0.2}$NiO$_2$ and poorly reduced non-superconducting Nd$_{0.8}$Sr$_{0.2}$NiO$_{3-x}$ thin films. The insets in (a) and (b) show the Ni L$_2$-edges. In (d) we show the H-dependence of the XMCD at the Ni L$_3$-edge for the uncapped NdNiO$_2$ sample, with the XMCD contribution due to the Ni$^{2+}$ fraction observed (inset). (e) H-dependence of the XMCD at the Ni L$_3$-edge for the capped NNO sample. (f) H-dependence of the Ni-$\mu_{spin}$ for different capped samples at 4\,K. Dotted lines are guide to the eyes.}
    \label{fig1}
\end{figure*}
All the samples were grown onto (001) SrTiO$_3$ (STO) single-crystals, with/without a STO capping-layer and the latter, when present, was only three-unit-cells thick ($\sim$1 nm). Details about the growth, structural and transport characterizations are provided elsewhere \cite{KriegerJAP}. The experiments were performed at DEIMOS and ID32 beamlines of the SOLEIL and ESRF Synchrotron facilities (France), respectively, which allowed us to verify that the results, obtained on different sample series prepared in nominal identical conditions, were reproducible (please refer to the Appendix \ref{appendix:rep}). The x-ray absorption spectra (XAS) were acquired in total electron yield (TEY) and in fluorescence yield (FY) modes, with beam size  of the order of 500$\times$800\,$\mu m^2$, in normal incidence (beam and magnetic field perpendicular to the NiO$_2$ planes) with circularly polarized light. The XMCD signal is proportional to the magnetic moment of a given ion (here Ni$^{1+}$, Ni$^{2+}$ and Nd$^{3+}$) projected along the x-ray beam and the external field. The Ni spin moment (Ni-$\mu_{spin}$) was determined by applying the sum-rules to the XMCD spectra \cite{PreziosiXMCD}. We used the nominal count of 1.0 and 1.2 holes per Ni-site in undoped and 20\% Sr-doped nickelates, respectively. 

First we show how the XMCD technique is able to single-out the intrinsic Ni$^{1+}$ (and Nd$^{3+}$) magnetic moments from the extrinsic contribution from non ideal-regions where Ni is in a Ni$^{2+}$ valence state. Hereof we show Ni-$L_{3,2}$ edge XAS and XMCD data on a series of STO-capped and uncapped NdNiO$_2$ thin films in both TEY and FY modes. In the TEY mode, the XAS and XMCD spectra have a major contribution from the top-most unit-cells, while in the FY-mode the entire film volume is equally probed. The upper-parts of Figures \ref{fig1}a,b show circular polarized Ni-$L_{3}$ edge XAS spectra (average of C-plus and C-minus polarized spectra), normalized to the L$_3$-edge intensity, acquired at H=9\,T and at T=4\,K in normal incidence geometry for both STO-capped and uncapped samples. The Ni-$L_{2}$ edge-spectra are shown in the related insets. 
The XAS of uncapped samples are broader than the ones measured in capped samples \cite{note}, due to an important contribution of the higher energy Ni$^{2+}$ feature akin to not fully reduced unit-cells, in agreement with other studies \cite{Raji2023,Parzyck2024}. 
\begin{figure*}[t!]
\centering
\includegraphics[width=1\textwidth]{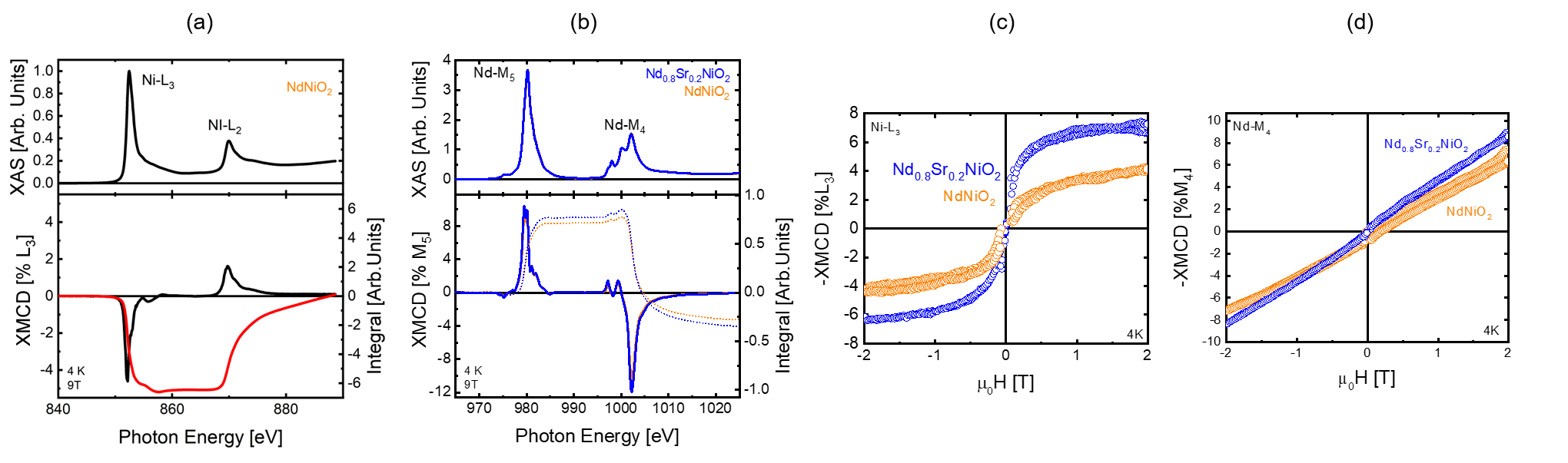}
\caption{NdNiO$_2$ and Nd$_{0.8}$Sr$_{0.2}$NiO$_2$ XAS (top-panel) and XMCD (bottom-panel) at the (a) Ni L$_{3,2}$- and (b) Nd M$_{5,4}$ -edges. The integrals of the XMCD spectra are shown as dotted-lines. Magnetic hysteresis at the (c) Ni L$_3$-edge and (d) Nd M$_{4}$ peaks, normalized to the pre-edge value for both samples.}
\label{fig2}
\end{figure*}
A comparison between TEY and FY XAS spectra shows that the Ni$^{2+}$ signal is mainly due to the first surface unit-cells, in agreement with scanning transmission electron microscopy and recent x-ray photoemission spectroscopy measurements \cite{Raji2023,MartandoXPS}. On the other hand, the TEY and FY XAS spectra of capped samples largely overlap and show only a Ni$^{1+}$ feature, with only small differences attributed to Ni$^{2+}$ located at the interface with the STO-capping-layer. These experimental evidences confirm that the STO-capping allows a better stabilization of the IL phase \cite{Lee2020}.

The differences between XAS spectra of capped and uncapped samples echoed in the XMCD data, shown in bottom part of Figures \ref{fig1}a,b. In particular, uncapped NdNiO$_{2}$ are characterized by a doublet structure in the TEY-XMCD spectra at 9\,T and 4\,K. This doublet almost disappears at low magnetic field as shown in Figure \ref{fig1}d, and it is absent in the FY-XMCD spectra (Fig. \ref{fig1}e), where a unique peak shows up. As highlighted by the vertical blue dotted lines, the two peaks in the TEY-XMCD doublet correspond to Ni$^{1+}$ and Ni$^{2+}$ features of the XAS spectra. It is worth noting that the TEY-XMCD peak intensity due to the Ni$^{2+}$ feature, normalized to the overall area of the XMCD at the Ni-$L_{3}$ edge, shows a linear H-dependence (see inset of Fig. \ref{fig1}d), suggesting a paramagnetic contribution from the incomplete reduction of the surface unit-cells. On the other hand, the XMCD intensity at the Ni$^{1+}$ photon energy dominates over the Ni$^{2+}$ contribution. 
In the case of the capped samples the FY-XMCD and TEY-XMCD spectra almost overlap, and are characterized by a unique and sharp peak mainly associated to the Ni$^{1+}$. A small Ni$^{2+}$ shoulder shows up only in the TEY-XMCD spectra and only at high magnetic fields, possibly due to the top-interface unit-cell. Finally, in order to confirm that the large XMCD in capped samples is related to Ni$^{1+}$-spins and it is intrinsic to our IL thin films, in Figure \ref{fig1}c we show additional XMCD data on an insulating, not fully reduced, capped Nd$_{0.8}$Sr$_{0.2}$NiO$_{3-x}$ sample, realized after a partial topotactic reduction process. As expected, here the XAS spectra are characterized by a more relevant Ni$^{2+}$ peak, a very large doublet in the XMCD signal and a factor five lower Ni$^{1+}$ XMCD even at the maximum magnetic field. 
Therefore, our XMCD results allow us to conclude that Nd-based IL-nickelates exhibit a relatively strong out-of-plane spin moment associated to the Ni$^{1+}$-spins in the NiO$_2$ planes, 
in qualitative agreement with Fowlie $et$ $al.$ \cite{Fowlie2022}. 

Now we address the magnetic properties of capped samples by discussing the XMCD data as a function of both H and T. In Figure \ref{fig1}f we show the Ni-$\mu_{spin}$ H-dependence in the 0.01-9\,T range at 4\,K estimated from the sum-rules applied to the Ni L$_{3,2}$-edges XMCD spectra of STO-capped NdNiO$_2$, Nd$_{0.8}$Sr$_{0.2}$NiO$_2$ and defective Nd$_{0.8}$Sr$_{0.2}$NiO$_{3-x}$ samples. Interestingly, the Ni-$\mu_{spin}$ at 9\,T is $\sim$0.12$\mu_B$/Ni and $\sim$0.25$\mu_B$/Ni for the undoped and superconducting samples, respectively, and $\sim$0.07$\mu_B$/Ni, in defective Nd$_{0.8}$Sr$_{0.2}$NiO$_{3-x}$. Moreover, all the samples exhibit a peculiar H-dependence characterized by a rapid increase below 1\,T followed by a slow linear increase at higher fields. 
 
The rather similar slope at high field for all samples and the stronger magnetism found in fully reduced NdNiO$_2$ samples characterized, according to XAS and XMCD by a minimal amount of Ni$^{2+}$ as compared to incompletely reduced Nd$_{0.8}$Sr$_{0.2}$NiO$_{3-x}$, exclude any major role of paramagnetism or superparamagnetism from Ni$^{2+}$-related defects (please refer to the dedicated section in the Appendix for further details).

In order to better understand the magnetism of our system, in Figures \ref{fig2}a,b we compare Ni $L_{3,2}$- and Nd M$_{5,4}$-edges XAS and XMCD spectra, acquired in normal incidence and 9\,T. The Nd M$_{5,4}$ XAS shows typical features of Nd$^{3+}$ $4f$ states \cite{NdXAS}, and a discussion of the sign of the Nd-XMCD is presented in the Appendix.   
\begin{figure*}[t]
\centering
\includegraphics[width=1\textwidth]{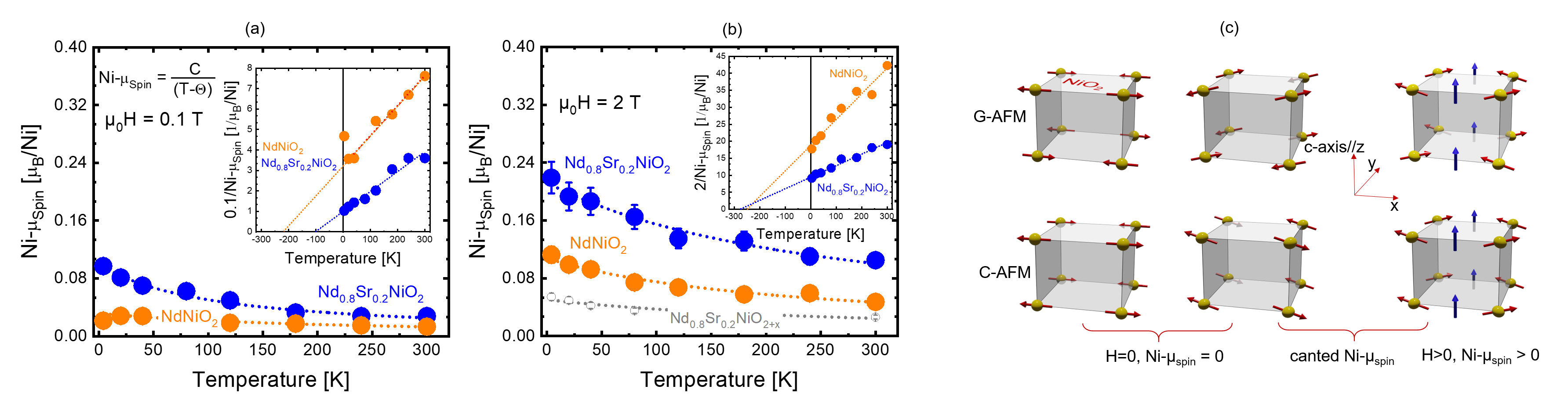}
\caption{Temperature dependence of the Ni-$\mu_{spin}$ for different samples acquired at (a) 0.1\,T and (b) 2\,T. Dotted lines are fits to the data by a Curie-Weiss-law. In the insets the corresponding inverse susceptibilities are shown. (c) (left) Sketches of in-plane G-type (top) and C-type (bottom) 2D-AFM orders within the NiO$_2$ planes; (middle) canted-AFM spin arrangement; (right) magnetic field induced out-of-plane spin-moment due to canting.}
\label{fig3}
\end{figure*}
In Figures \ref{fig2}c,d we show the magnetization hysteresis loops, obtained by measuring the XMCD at the Ni $L_3$ ($\sim$852\,eV) and Nd M$_4$ ($\sim$981\,eV) edges. The data are overall proportional to the Ni$^{1+}$ and Nd$^{3+}$ out-of-plane magnetic moments. The Ni $L_3$ XMCD loops show apparently no clear hysteresis, and the data well agree with the H-dependence of the Ni-$\mu_{spin}$ determined from the sum-rules, again suggesting that the Ni$^{1+}$ intrinsic magnetization is characterized by a very steep increase below 1\,T followed by an H-linear slow increase for both samples. Moreover, the data confirm that the Ni-$\mu_{spin}$ is substantially larger for the doped superconducting sample. The Nd M$_4$ XMCD hysteresis loop, on the other hand, shows a paramagnetic signal to a large extent uncorrelated to the Ni-$\mu_{spin}$ H-dependence, although a very small coupling at low field cannot be excluded since the XMCD signals show some deviation from a purely linear trend in this region of fields.

In order to get further insights about the magnetic correlations that are at play, we show in Figure \ref{fig3} the Ni-$\mu_{spin}$ and inverse susceptibility (1/$\chi$ $\sim$ H/Ni-$\mu_{spin}$) temperature dependencies in undoped and Sr-doped (superconducting) samples at field values of 0.1\,T and 2\,T. The latter were chosen to get insights from the two different regions of the Ni-$\mu_{spin}$ H-dependence ($cf.$ Fig. \ref{fig1}f). At 2\,T in both samples, the Ni-$\mu_{spin}$ increases with decreasing temperature, obeying with a good accuracy to a Curie–Weiss-law in the entire range, as quantified in the fitting procedure fully reported in the Appendix. 
In weak magnetic fields (0.1\,T), on the other hand, we observe a downturn of the Ni-$\mu_{spin}$ in undoped NdNiO$_2$, not observed in doped sample and disappearing at high field, because it is masked by a stronger field-induced moment. At the same time, the inverse susceptibilities (insets of Figs. \ref{fig3}a,b), show Curie-Weiss-like linear T-dependence that extrapolate to negative temperature values, at odds with (super)paramagnetism or ferromagnetism. Both results, $i.e.$ the anomaly in the temperature dependence of the Ni-$\mu_{spin}$ and the intercept to negative temperatures of the inverse susceptibilities data calls for dominant AFM correlations within the NiO$_{2}$ planes.

To explain the origin of the large field-induced Ni-$\mu_{spin}$ we consider a canted (G-type or C-type) AFM spin ordering as sketched in Figure \ref{fig3}c. Since with XMCD we cannot establish if a long-range static AFM order is present in our samples, those sketches have to be considered as a cartoon of an antiferromagnetically correlated spin system where both static (and ordered) arrangement, as well as fluctuations of the Ni-spins, can be realized. The canting of the Ni$^{1+}$-spins is, for sake of simplicity, represented in the $xz$ plane. At zero field the out-of-plane spin moment is null. Upon H-increasing, the intrinsic magnetic instability of the Ni-spin-sublattice, associated to a weak out-of-plane coupling, allows a partial spin-flip of the (canted) spins, with consequent alignment along the field and the emergence of a non zero Ni-$\mu_{spin}$ out-of-plane component (indicated by the blue arrows). In particular, the canted component reaches its maximum with the applied field of approximately 1\,T. The canted spin-moment is locked to the in-plane antiferromagnetically correlated spin. At 0.1\,T, the temperature dependence of the magnetic moment reflects the in-plane AFM order or correlations in NdNiO$_2$, giving rise to a decrease of the Ni-$\mu_{spin}$ out-of-plane component below 50\,K, as shown in Figure \ref{fig3}a. The slow linear increase above 1\,T is therefore due to a very small and slow increase of the spin-canting angle, because the in-plane AFM coupling is very strong. In Sr-doped samples the larger Ni-$\mu_{spin}$ out-of-plane component ($cf.$ to Fig. \ref{fig1}f), can be associated to a larger canting, since the AFM correlation is weaker in the doped compound \cite{LuMagnon,rosa2024spin}.

\begin{figure}[t!]
\centering
\includegraphics[width=0.4\textwidth]{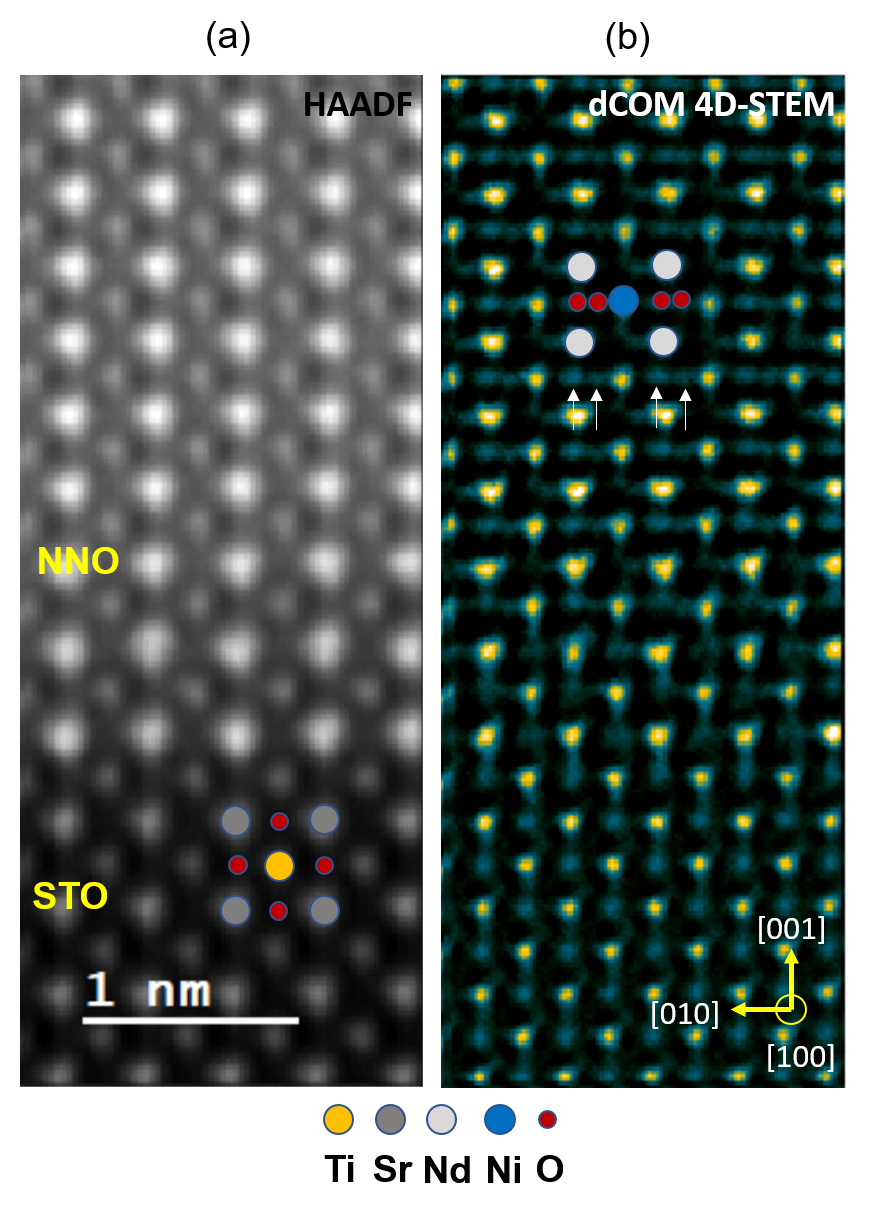}
\caption{(a) High-angle annular dark-field imaging (HAADF) of capped NNO sample acquired at 300\,K combined to a (b) dCOM 4D-STEM imaging showing the NiO$_2$ planes. The presence of oxygen-ions along the Ni-O-Ni bond with different center of mass positions are indicated by the arrows.}
\label{fig4}
\end{figure}

While the suggested model can simultaneously explain the presence of in-plane AFM correlations and the relative large H-induced spin-moment, the origin of the spin-canting has to be settled. Our XMCD data clearly demonstrate that the measured magnetic moment should be attributed to Ni$^{1+}$ in the NiO$_2$ planes. However, perfectly arranged square-planar NiO$_2$ would not give rise to an out-of-plane spin-canting due to the lack of buckling as in the case of Sr$_{2}$CuO$_{2}$Cl$_{2}$ oxychloride cuprates. On the other hand, a canted spin-moment was observed in the La$_{2-x}$Sr$_x$CuO$_4$ and R$_1$Ba$_2$Cu$_3$O$_7$ families, and attributed to the Dzyaloshinskii–Moriya interaction (DMI) \cite{DeLuca2010} (and references therein), taking place due to non-perfectly square-planar Cu-O-Cu bonds, thus to a lowering of the symmetry. In doped and superconducting La$_{2-x}$Sr$_x$CuO$_4$ and R$_1$Ba$_2$Cu$_3$O$_7$, the weak-ferromagnetic out-of-plane component has a pure paramagnetic temperature and magnetic field dependence, while in undoped La$_{2}$CuO$_4$ the temperature dependence reflects the in-plane AFM long-range order. In analogy with cuprates, we tried to identify any (structural) sources of the spin-canting in our NdNiO$_{2}$ by using divergence of the Center of Mass (dCoM) four-dimensional (4D) scanning transmission electron microscopy (STEM-dCOM). In 4D-STEM, we collect the whole electron diffraction, that enables us to quantify the shift happened to the center of mass (COM) of the electron diffraction at each pixel in real space as the probe scans the sample. The momentum transfer to the electron beam on interacting with the sample is inferred from this COM measurements. This also becomes a qualitative representation of the in-plane electric field, with it being negatively proportional to the momentum transfer. Thus, the divergence of in-plane electric field, approximates to a projected charge-density image, with extrema located at the atomic sites. This is called the divergence of the center of mass (dCOM) technique. It can enable good phase contrast imaging of lighter elements like oxygen, at a sub-angstrom resolution \cite{Hachtel2018,Alex4D}. In particular, the 4D-STEM dCOM has been fundamental in the discovery of a novel nickelate phase emerging from three-dimensional oxygen vacancy ordering \cite{Raji2024NewPhase}. Figures \ref{fig4}a,b show high-angle annular dark field STEM and 4D-STEM dCOM map images, respectively, along the [100] cubic axis of STO, for an undoped capped NdNiO$_{2}$ film. The NiO$_4$ square-planar coordination is altered and presents an apparent oxygen-doubling with a zig-zag pattern, as indicated by the white arrows in Figure \ref{fig4}b. The presence of an additional (faded) oxygen along the Ni-O-Ni bond suggests that the oxygen ions are not exactly at the center. From the data, we cannot exactly identify the source of this non-ideal planar-configuration, which might originate from tilt/rotation of the NiO$_2$ bonds and/or to a cation off-centering. Nonetheless, the overall symmetry of the structure appears lowered by these structural distortions, and this can explain a sizable DMI and a non-collinear Ni$^{1+}$-spin arrangement \cite{DMIreview}.

In conclusion, with our elemental-sensitive XMCD experiments at the Ni $L_{3,2}$- and Nd $M_{5,4}$-edges we could demonstrate that the magnetic properties of Nd-based IL-nickelate parent compound and superconducting thin films are related to a canting of AFM correlated Ni$^{1+}$-spin, ruling out the contribution from Ni$^{2+}$ and of extended defects to the measured magnetism. Moreover, we showed that the paramagnetic Nd$^{3+}$ $4f$ states do not affect the overall Ni-$\mu_{spin}$ H-dependence. The spin-canting is attributed to non-perfectly square-planar NiO$_2$ bonding, as evidenced by 4D-STEM dCOM measurements, which lowering the symmetry can be responsible for a sizable DMI among the Ni$^{1+}$ spin. The presence of a large out-of-plane spin moment, in a range of magnetic fields where superconductivity is still very robust, has consequences on the superconducting properties. In particular, the very large upper critical field, substantially higher than the Pauli-limit according to recent studies \cite{Wang2021}, could be possibly linked to the magnetism of the system.
\section{Appendix}
Here we report about some aspect of our work mainly dealing with the reproducibility of the data shown in Figure \ref{fig1}, and the analysis followed to exclude any possible paramagnetism and/or superparamagnetism character of the measured magnetism for both reduced and partially reduced nickelate samples.  
\setcounter{figure}{0}
\renewcommand{\thefigure}{A\arabic{figure}}

\subsection{XAS and XMCD measurements: Reproducibility}\label{appendix:rep}
\begin{figure}[h]
\centering
\includegraphics[width=0.35\textwidth]{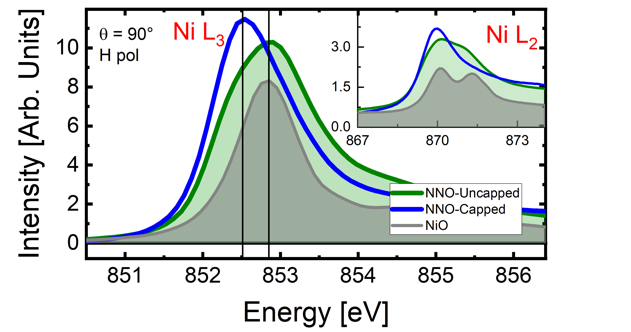}
\caption{TEY-XAS for capped and uncapped NNO samples. A comparison with NiO for the reference of Ni$^{2+}$ is also reported. Re-adapted from \cite{LeonardoThesis}.}
\label{figA1}
\end{figure}

The differences in the fraction of Ni$^{2+}$ in capped and uncapped NNO samples, and the differences in FY and TEY data in uncapped samples, are in full agreement with other studies, in particular extensive x-ray photoemission spectroscopy \cite{Raji2023,rath2024scanning}. These data were also obtained on a different series of undoped samples (prepared in the same conditions) with and without a capping layer. Similar XAS data were already published in Refs. \cite{Raji2023,kriegerPRL} on capped and uncapped samples on other sample series. All the data are fully consistent with those presented in Fig. \ref{fig1}.
As an example, in Figure \ref{figA1} we show additional TEY-XAS data acquired on one of our first generation of NdNiO$_{2}$ samples prepared with and without a capping layer. It is worth noting that, as in the TEY-XAS data presented in the main text, the Ni$^{2+}$ feature in the XAS is more pronounced in the uncapped samples, and since in TEY we are more sensitive to the top unit cells, these data compare well with the results reported in the main text about uncapped NdNiO$_{2}$, where the presence of Ni$^{2+}$, besides the Ni$^{1+}$ feature, is clear. 

\begin{figure}[h]
\centering
\includegraphics[width=0.35\textwidth]{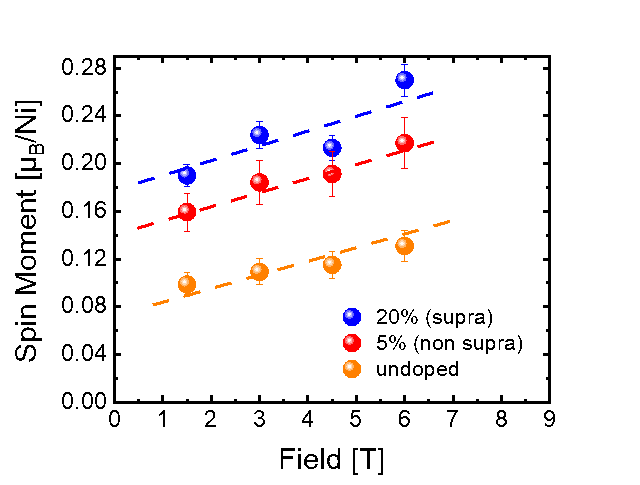}
\caption{H-dependence of the Ni$^{1+}$-spin for a superconducting (20\% at. Sr), non-superconducting (5\% at. Sr) and undoped NdNiO$_2$ sample. }
\label{figA2}
\end{figure}

Figure \ref{figA2} shows XMCD data as a function of the magnetic field for a first series of Sr-doped and undoped NdNiO$_{2}$ samples. These data were acquired at magnetic fields above 1 Tesla. We found a linear trend of the Ni$^{1+}$-spin moment, and values fully comparable to the ones reported in Figure \ref{fig1}f of the main text. This demonstrates that the results presented in this work are fully reproducible.

\subsection{Supplementary analysis of the Ni-$\mu_{spin}$ as a function of the magnetic field.}

\begin{figure}[h]
\centering
\includegraphics[width=0.5\textwidth]{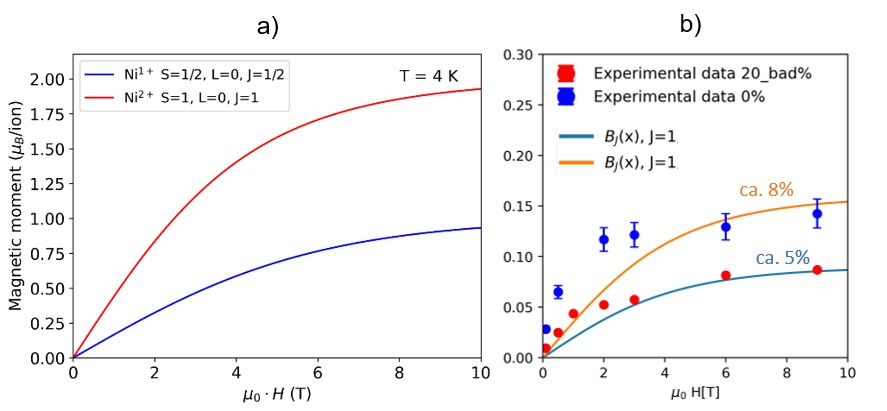}
\caption{(a) Brillouin simulation for Ni$^{1+}$ (1 $\mu_B$) and Ni$^{2+}$ (2\,$\mu_B$) magnetic moments at 4\,K. b) Ni$^{2+}$ (J=1, 2\,$\mu_B$) Brillouin function modified to saturate to the experimental data at the highest H.}
\label{figA3}
\end{figure}

\begin{figure}[h]
\centering
\includegraphics[width=0.4\textwidth]{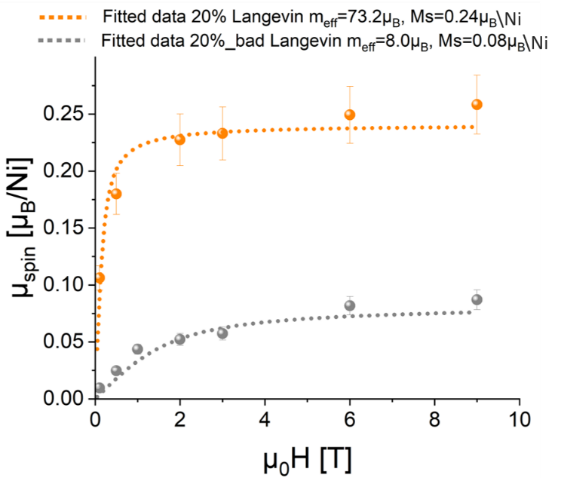}
\caption{Results from a Langevin fitting of the M(H) curves presented in Fig. \ref{fig1}f with m$_{eff}$ being the effective moments of the clusters and Ms the magnetic moment saturation values.}
\label{figA4}
\end{figure}

In order to exclude trivial paramagnetism from Ni$^{1+}$ and Ni$^{2+}$ magnetic moments, we show in Figure \ref{figA3} the calculated magnetic field dependence at 4\,K in the 0-10 Tesla range obtained using a simple Brillouin function (as expected for paramagnetism). The experimental data of Figure \ref{fig1}f, characterized by a rapid magnetic moment increase at low field and a non-saturating H-linear behaviour, cannot be reproduced by pure (non-interacting) paramagnetism. In particular, the expected nominal Ni magnetic moments of 1 $\mu_B$/Ni (2 $\mu_B$/Ni) for the Ni$^{1+}$-3d$^9$ (Ni$^{2+}$-3d$^8$) electronic configuration, according to the Brillouin function (cf. Figure \ref{figA3}), should align rather smoothly at fields much higher than 10 Tesla. Moreover, the rapid low-field increase cannot be explained by pure paramagnetism, nor the linear slow increase above 1 Tesla.
Another possible mechanism is super-paramagnetism from unidentified defect clusters, alike Ni oxides recently proposed from scanning squid microscopy studies \cite{Shi2024}, or other defects where Ni is in 2+ (or 3+) oxidation state. 
In Figure \ref{figA4} we show a Langevin fit of the data on reduced (and SC) capped 20\%-NSNO sample and incompletely reduced (fully insulating) capped 20\%-NSNO. Although capturing the initial rapid increase of the magnetic moment in both cases, the cluster sizes obtained from the analysis are far beyond any reasonable value, especially in the case of the fully reduced superconducting NSNO. Indeed, we obtain an effective magnetic moment (m$_{eff}$) of 73.2\,$\mu_B$ in (superconducting) 20\%-doped, capped NSNO. Assuming a Ni$^{2+}$ valence in these defective regions, we get about 25\% of the sample volume characterized by (fairly large) defective clusters. We did not find from our STEM data any NiO$_x$ cluster or defective NSNO regions at such a density to explain superparamagnetism. Moreover, the XAS data cannot be reproduced assuming a fraction of Ni$^{2+}$ in the SC sample of 12.5\% \cite{kriegerPRL}. 

More importantly, from the Langevin fitting it emerges that the cluster sizes in the defective 20\% NSNO (not fully reduced) are much smaller (about 8\,$\mu_B$) and the fraction of regions with Ni$^{2+}$ valence in the sample should be at most 5\% or less. On the other hand, the XAS data clearly show that the amount of Ni$^{2+}$ in the defective 20\% NSNO is much larger than in the fully reduced SC NSNO, which further demonstrates that the magnetism observed cannot be due to super-paramagnetism from defect-clusters. 
Finally, superparamagnetism cannot explain the slow linear increase of the magnetic moment observed in all samples. In particular, the slope of this linear magnetic field dependence is only marginally dependent on the sample properties, and in particular on the amount of Ni$^{2+}$. This result again reinforces our analysis, and in particular it demonstrates that the magnetism cannot be attributed to unidentified defects or to pure paramagnetism. 

\subsection{Supplementary analysis of the Nd M$_{5,4}$-edges XMCD}

To investigate any possible cross-talk from the magnetism of the Nd-$4f$ states, we performed also XMCD measurements at the Nd $M_{5,4}$-edges, as shown in Figure \ref{fig2}b where we report the total-XAS and related XMCD acquired at 4\,K and 9\,T for both samples. 
The total-XAS shows typical features of a stable Nd$^{3+}$ valence state as already reported in literature \cite{NdXAS,R_XMCD} with a $J=9/2$, $S=3/2$ and $L=6$ ground state as dictated by Hund's rule with an expected (free ion) total magnetic moment of ca. 3.6\,$\mu_B$/Nd at which both spin and orbital components concur. The application of sum-rules in the case of rare-earth is not straightforward, and we have fully followed the procedure analysis reported in Ref. \cite{NdXAS}. The value obtained for the spin (Nd-$\mu_{spin}$) and orbital (Nd-$\mu_l$) magnetic moments are shown in Table \ref{table1}. While the opposite sign between the spin and orbital magnetic moments is expected from the Hund's rules for a less than half-filled $4f$ shell, the relatively low values measured especially for the orbital component might originate from a quenching process due to the specific crystalline field of the IL structure. Another striking difference is stemming from the values of both spin and orbital magnetic moments of each samples, which in principle were expected to be the same. Within the experimental error, the superconducting sample has larger values which might call for some key differences in the values of the correction factor and/or the quadrupole moments due to a different Nd environment as imposed by the Sr-doping. More insights could be obtained from an accurate analysis of the XAS and Nd-XMCD profiles which is, nevertheless, beyond the scope of this work. The relatively small values for the Nd-$\mu_l$ cause the Nd-XMCD sign to be entirely determined by the sign of the Nd-$\mu_{spin}$, which is completely the opposite situation found in literature for several others Nd-based compounds \cite{NdXAS,R_XMCD}. To understand at which extent the Nd magnetic moments enter in relation with the Ni ones we performed also hysteresis measurements at the Nd-M$_4$ edge, and we found that the Nd-$\mu_l$ are parallel to the Ni-$\mu_{spin}$, which on the contrary are aligned antiparallel to the Nd-spin moments.

\begin{table}[h!]
   \caption{Spin and orbital magnetic moments calculated from the XMCD at the Nd $M_{5,4}$-edges. Nd-$\mu_{spin}$ and Nd-$\mu_l$ are non-parallelly aligned as expected for a less than half-filled $4f$ shell according to Hund's rules.
    }
    \centering
    \begin{tabular}{ccc}
    \hline 
    \rowcolor[HTML]{EFEFEF}
    Sample & $\mu_l~[\mu_B/Nd]$ & $\mu_{spin}~[\mu_B/Nd]$ \\ \hline
     \makecell{Superconducting \\ $\mathrm{Nd}_{0.8}\mathrm{Sr}_{0.2}\mathrm{NiO}_2$} & 0.44$\pm$0.02 & -0.62$\pm$0.02 \\ \hline
    \makecell{Insulating \\ $\mathrm{Nd}\mathrm{NiO}_2$} & 0.35$\pm$0.02 & -0.50$\pm$0.02   \\ \hline
			 \end{tabular}
    \label{table1}
\end{table}

\subsection{Supplementary analysis of the Ni-$\mu_{spin}$ magnetic moment as a function of the temperature}

We performed a fitting of the Ni-$\mu_{spin}$(T) data using the Curie-Weiss law:  Ni-$\mu_{spin}$/B=$\chi$= C/(T-$\theta$), with C in Tesla$^{-1}$\,$\mu_B$\,K. The Curie-Weiss law can fit the data very well, and very interestingly the $\theta$ intercept corresponds to negative temperature values. In Tables \ref{table2},\ref{table3} we report the fitting parameters for the two samples analysed for both magnetic field values. Note the negative values of the $\theta$ parameters indicate the dominant contribution of AFM correlations. 

\begin{table}[h!]
   \caption{ Curie-Weiss fitting at the magnetic field of 2 Tesla.}
    \centering
    \begin{tabular}{ccc}
    \hline 
    \rowcolor[HTML]{EFEFEF}
    \makecell{Fitting formula \\ C/(T-$\theta$)} & \makecell{Superconducting \\ $\mathrm{Nd}_{0.8}\mathrm{Sr}_{0.2}\mathrm{NiO}_2$}  & \makecell{Insulating \\ $\mathrm{Nd}\mathrm{NiO}_2$} \\ \hline
     \makecell{C \\ (Tesla$^{-1}$\,$\mu_B$\,K)} & 27.50$\pm$2.12  & 11.66$\pm$1.12 \\ \hline
        \makecell{$\theta$ \\(K)} & -255.69$\pm$23.81 & -211.74$\pm$24.80   \\ \hline
        Adj. R-Square & 0.97991 & 0.95889 \\ \hline
			 \end{tabular}
    \label{table2}
\end{table}

\begin{table}[h!]
   \caption{ Curie-Weiss fitting at the magnetic field of 0.1 Tesla.}
    \centering
    \begin{tabular}{ccc}
    \hline 
    \rowcolor[HTML]{EFEFEF}
    \makecell{Fitting formula \\ C/(T-$\theta$)} & \makecell{Superconducting \\ $\mathrm{Nd}_{0.8}\mathrm{Sr}_{0.2}\mathrm{NiO}_2$}  & \makecell{Insulating \\ $\mathrm{Nd}\mathrm{NiO}_2$} \\ \hline
     \makecell{C \\ (Tesla$^{-1}$\,$\mu_B$\,K)} & 10.14$\pm$0.73  & 6.55$\pm$0.70 \\ \hline
        \makecell{$\theta$ \\(K)} & -101.20$\pm$12.03 & -205.83$\pm$37.13   \\ \hline
        Adj. R-Square & 0.964738 & 0.944167 \\ \hline
			 \end{tabular}
    \label{table3}
\end{table}

\section{Acknowledgments}
This work was funded by the French National Research Agency (ANR) through the ANR-JCJC FOXIES ANR-21-CE08-0021. This work was also done as part of the Interdisciplinary Thematic Institute QMat, ITI 2021 2028 program of the University of Strasbourg, CNRS and Inserm, and supported by IdEx Unistra (ANR 10 IDEX 0002), and by SFRI STRAT’US project (ANR 20 SFRI 0012) and EUR QMAT ANR-17-EURE-0024 under the framework of the French Investments for the Future Program. The synchrotron experiments were performed at ESRF and SOLEIL Synchrotron facilities in France under proposal numbers SC-5296 and 20200718, respectively. D.P. acknowledges fruitful discussions with Gabriele de Luca regarding the Nd sum-rules, and Guillaume Rogez about magnetism. André Thiaville is also acknowledged for discussions over the possible presence of DMI in our NdNiO$_2$ thin films. 

\section{Bibliography}
\bibliography{nsno.bib}
\end{document}